\documentclass[times, 10pt, twocolumn, latex8]{article} 
\usepackage[latin1]{inputenc}
\usepackage[francais]{babel}
\usepackage{latex8}
\usepackage{times}
\usepackage{graphicx}
\usepackage{amsmath}
\usepackage{subfig}
\usepackage[ps2pdf=true,pagebackref=true]{hyperref}
\hypersetup{
    pdftitle =
        Activities of Daily Living Indexing by Hierarchical HMM for Dementia Diagnostics,
    pdfauthor =
        Svebor Karaman,
}

\begin{document}

\title{Activities of Daily Living Indexing by Hierarchical HMM for Dementia Diagnostics}

\author{Svebor Karaman, Jenny Benois-Pineau\\
LaBRI, Universit\'e de Bordeaux\\ 351 Cours de la Lib\'eration 
33405 Talence cedex, France\\ karaman@labri.fr, jenny.benois@labri.fr\\
\and
R\'emi M\'egret\\
IMS, Universit\'e de Bordeaux\\ 351 Cours de la Lib\'eration 
33405 Talence cedex, France\\ megret@enseirb-matmeca.fr\\
\and
Julien Pinquier\\
IRIT, Universit\'e de Toulouse\\ 118 route de Narbonne 
31062 Toulouse cedex 9, France\\ pinquier@irit.fr\\
\and
Yann Ga\"estel, Jean-Fran\c{c}ois Dartigues\\
Universit\'e Victor Segalen Bordeaux 2\\ 146 rue L\'eo Saignat
33076 Bordeaux Cedex, France\\ yann.gaestel@isped.u-bordeaux2.fr, dartigues@isped.u-bordeaux2.fr}

\maketitle

\begin{abstract}
This paper presents a method for indexing human activities in videos captured from a wearable camera being worn by patients, for studies of progression of the dementia diseases. Our method aims to produce indexes to facilitate the navigation throughout the individual video recordings, which could help doctors search for early signs of the disease in the activities of daily living. The recorded videos have strong motion and sharp lighting changes, inducing noise for the analysis. The proposed approach is based on a two steps analysis. First, we propose a new approach to segment this type of video, based on apparent motion. Each segment is characterized by two original motion descriptors, as well as color, and audio descriptors. Second, a Hidden-Markov Model formulation is used to merge the multimodal audio and video features, and classify the test segments. Experiments show the good properties of the approach on real data.
\end{abstract}

\Section{Introduction}
Our society is aging, with a longer lifetime expectancy come new challenges, one of them is to help the elderly keep their autonomy as long as possible. The aging diseases result in a loss of autonomy. Dementia diseases of the elderly have a strong impact on activities of daily living (ADL). Medical studies \cite{Dartigues2005} have shown that early signs of diseases such as Alzheimer can be identified up to ten years before the actual diagnostics. Therefore the analysis of possible lack of autonomy in the ADL is essential to establish the diagnostics as soon as possible and give all the help the patient and his relatives may need to deal with the disease. Until now, the medical diagnostics are most of the time based on an interview of the patient and the relatives. The answers to a survey about how well the patient executes ADL allow an evaluation of the patient's situation. The main issue with this methodology is the lack of objectivity of the patient and his entourage.\\
The best way to determine the autonomy of one patient is to analyze his ability to execute the ADL in his own environment. However, it can be complicated for a doctor to come and watch the patient doing these ADL, as this would be a very time consuming task. It can be interesting to record the patient doing ADL with a camera. This is the idea of the project IMMED\footnote{\texttt{http://immed.labri.fr/}} (Indexing Multimedia Data from wearable sensors for Diagnostics and treatment of Dementia): to use a wearable camera to record the ADL \cite{Megret2008} with a view as close as possible to the patient's view. Wearable cameras have been used in the SenseCam project \cite{Hodges2006} for an automatic creation of a visual diary. In the WearCam project \cite{Picardi2007} a camera is strapped on the head of young children, the collected data are then analyzed in order to diagnose autism. In our context, the video brings an objective view of what the patient is doing and may permit the doctor to give a better evaluation of the patient's situation. The doctor cannot visit all patients and wait while they are doing the ADL.  Therefore, the videos will be recorded while the patients are visited by a medical assistant. The medical assistant will then upload the video to a server, and the automatic analysis will be fulfilled to index the ADL. The doctor will use these indexes for an easier navigation through the video and will then use the activities as a source of information to refine the diagnosis.\\
In our previous works \cite{Megret2008} and actual demonstrations \cite{VideoACMMM2010} we have already largely described the acquisition set-up and the general framework of the project. In this paper we focus on the core of indexing method we propose, the motion based segmentation and the HMM. HMMs have been successfully applied to audio analysis \cite{Rabiner1989} and in molecular biology \cite{Sjolander2003}. The application of HMMs to videos can be whether at a low level, for cut detection \cite{Borezcky1996}, or at a higher level aiming to reveal the structure according to a previously defined grammar, such as the events of a tennis match \cite{Kijak2003}. In the HMM both are important: the adequate description space (observations) on one hand, the state set and the connectivity expressed by a state transition matrix on the other.\\
The contributions of this paper are in proposing adequate HMM structure and also use of heterogeneous multimodal descriptor space, which has never been done before, for the best of our knowledge, in wearable video analysis. Hence, in section~\ref{Section.VideoAcquisitionSetup} we will briefly describe the acquisition set-up as it is today after several adjustments brought by medical practitioners and information scientists. We also qualify the very specific video acquired with this device. Section~\ref{Section.MotionAnalysis} describes the motion analysis and the motion based segmentation and section~\ref{Section.DescriptionSpace} presents the definition of the description spaces. The structure of the HMM we designed is explained in section~\ref{Section.HMM}. Experiments and results are shown in section~\ref{Section.Experiments} and conclusions and perspectives in section~\ref{Section.ConclusionsAndPerspectives}.

\Section{Video acquisition setup}
\label{Section.VideoAcquisitionSetup}

\SubSection{The device}
The video acquisition device should be easy to put on, should stay in the same position even when the patient moves strongly and bring as less discomfort as possible to an aged patient. Regarding these constraints, a vest was adapted to be the support of the camera. The camera is fixed with hook-and-loops fasteners which allow the camera's position to be adapted to the patient's morphology.

\begin{figure}[h]
 \centering
  \subfloat[Motion blur due to strong motion.]{\label{Fig.Frames.MotionBlur}\includegraphics[width=2.78cm]{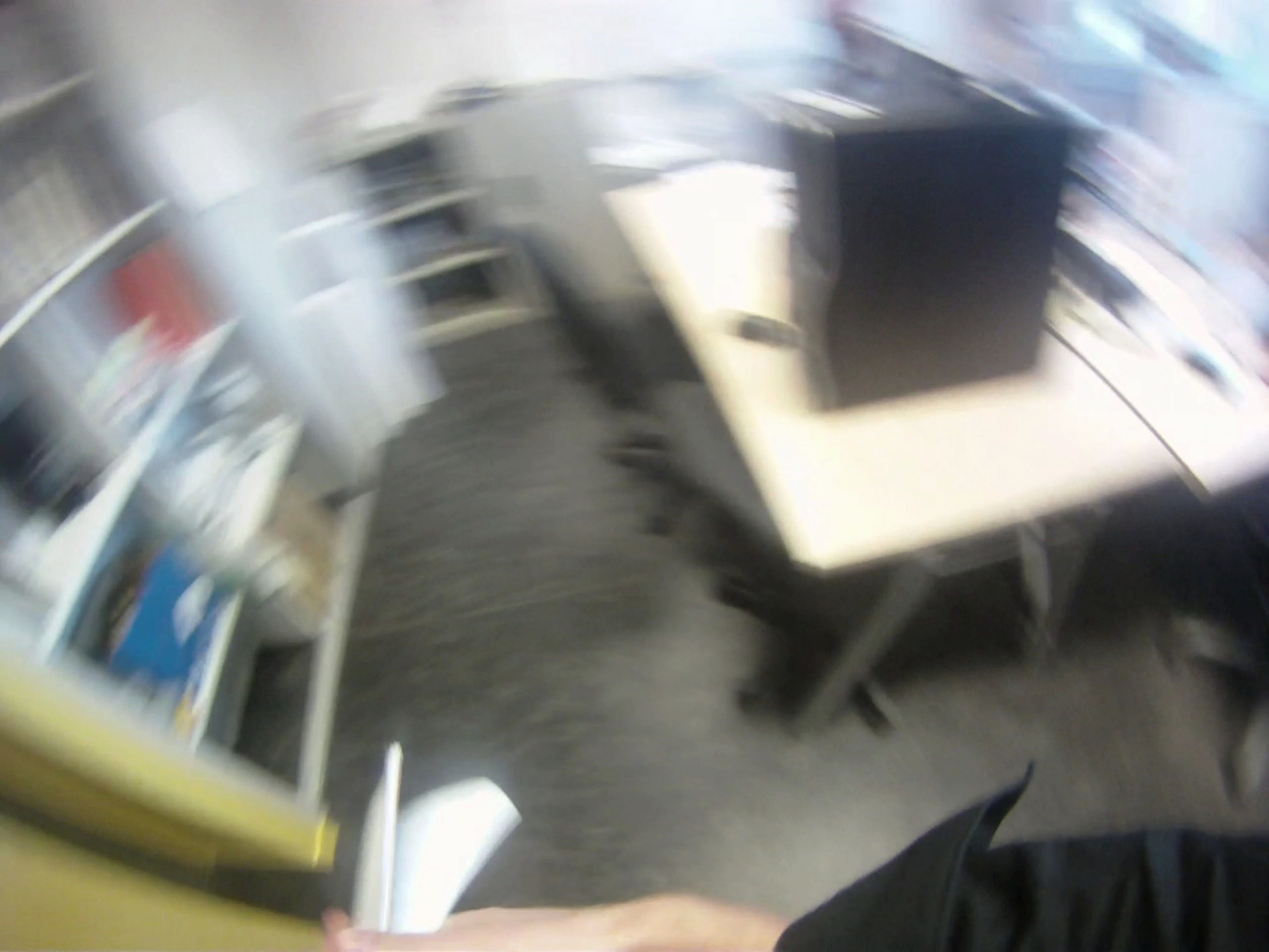}}
  \subfloat[Low lighting while in dark environment.]{\label{Fig.Frames.LowLighting}	\includegraphics[width=2.78cm]{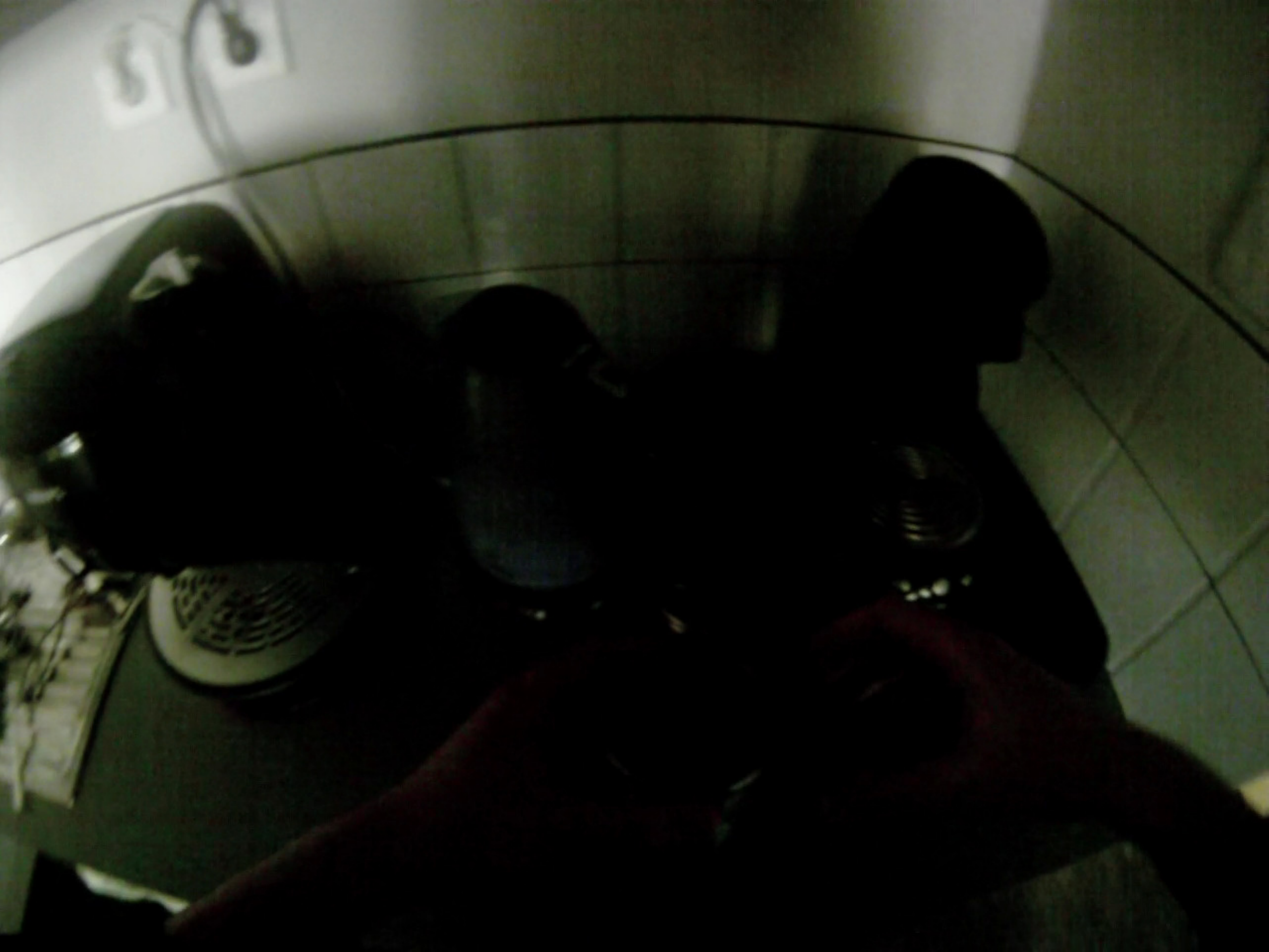}}
  \subfloat[High lighting while facing a window.]{\label{Fig.Frames.HighLighting}\includegraphics[width=2.78cm]{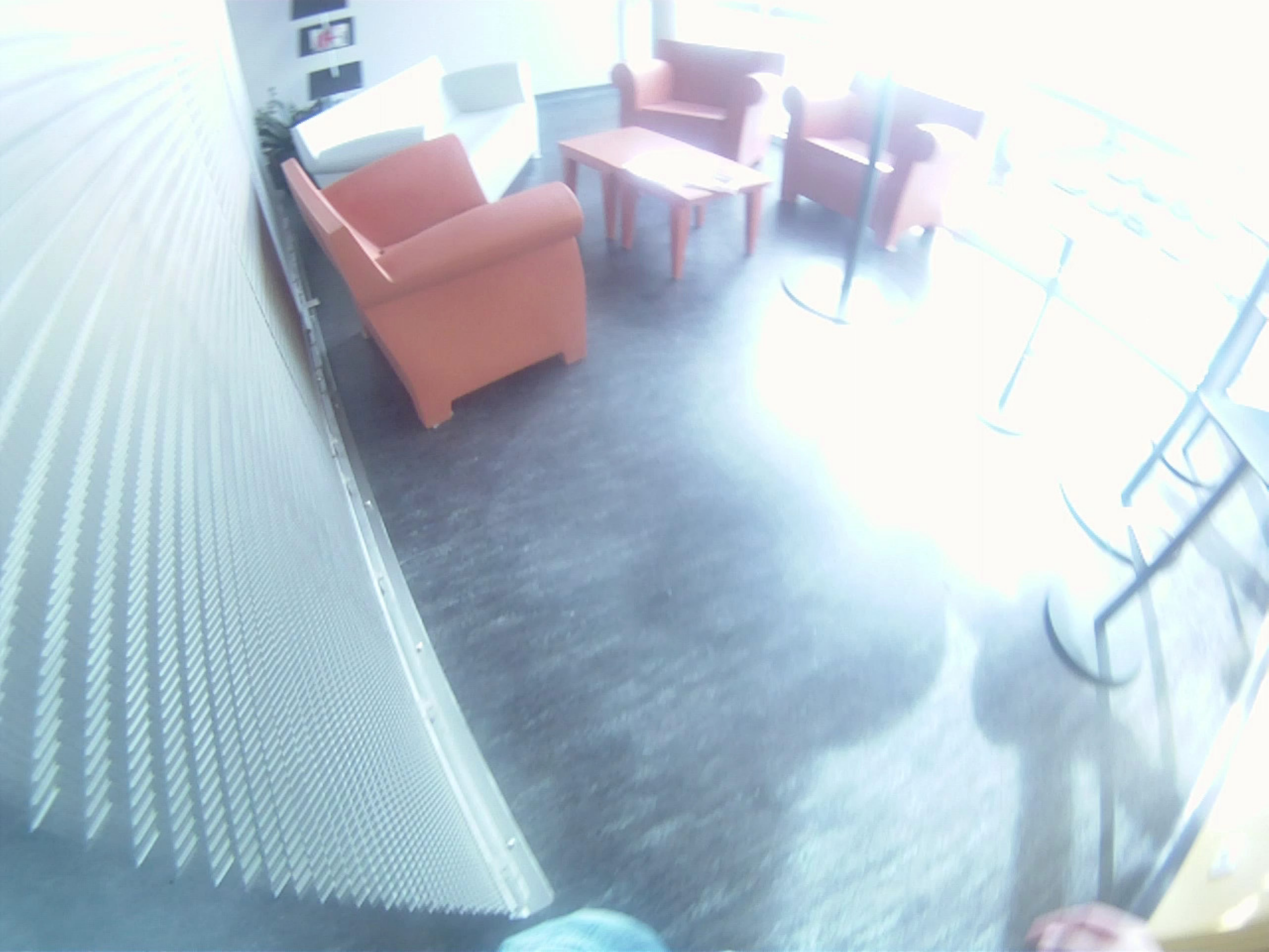}}
  \caption{Example of frames acquired with wearable camera.}
  \label{Fig.Frames}
  \centering
\end{figure}

\SubSection{Video characteristics}
The videos obtained from wearable cameras are quite different from the standard edited videos (having clean motion and cut into shots) which are usually subject to video indexing methods. Here, the video is recorded as a long sequence where the motion is really strong since the camera follows the ego-motion of the patient. This strong motion may produce blur in frames, figure~\ref{Fig.Frames}a. Moreover, the patient may face a light source, leading to sharp luminosity changes, figure~\ref{Fig.Frames}b and~\ref{Fig.Frames}c. The camera has a wide angle objective in order to capture a large part of the patient's environment.

\noindent

\section{Motion analysis for the design of description space}
\label{Section.MotionAnalysis}
In contrast to the work in \cite{Hodges2006} where the description space is based on a key-framing of the video, our goal is to use motion of the patient as one of the features. This choice corresponds to the need to distinguish between various activities of a patient which are naturally static (e.g. reading) and dynamic (e.g. hoovering).

\subsection{Global motion estimation}
Since the camera is worn by a person the global motion observed in image plane can be called the ego-motion. We model the ego motion by the first order complete affine model and estimate it with a robust weighted least squares by the method we reported in \cite{Benois2005}. The parameters of (\ref{Eq.Motion}) are computed from the motion vectors extracted from the compressed video stream.

\begin{equation}
\begin{pmatrix} dx_{i} \\ dy_{i} \end{pmatrix}  = \begin{pmatrix} a_1 \\ a_4 \end{pmatrix}  +  \begin{pmatrix} a_2 & a_3 \\ a_5 & a_6 \end{pmatrix}  \begin{pmatrix} x_i \\ y_i \end{pmatrix}
\label{Eq.Motion}
\end{equation}
Eq.~\ref{Eq.Motion}: Motion compensation vector, ($x_i$, $y_i$) being the coordinates
of a block center. 

\subsection{Motion-based segmentation}

In order to establish a minimal unity of analysis which may be considered as an equivalent to shots in our long sequence videos, we designed a motion based segmentation of the video. The objective is to segment the video into different viewpoints that the patient provides by moving throughout his home.

\subsubsection{Corner trajectories}
\label{Section.CornerTrajectories}

\begin{figure}[h]
 \centering
  \subfloat[Corner trajectories while the person is static. ]{\label{Fig.Corners.Static}\includegraphics[width=4.1cm]{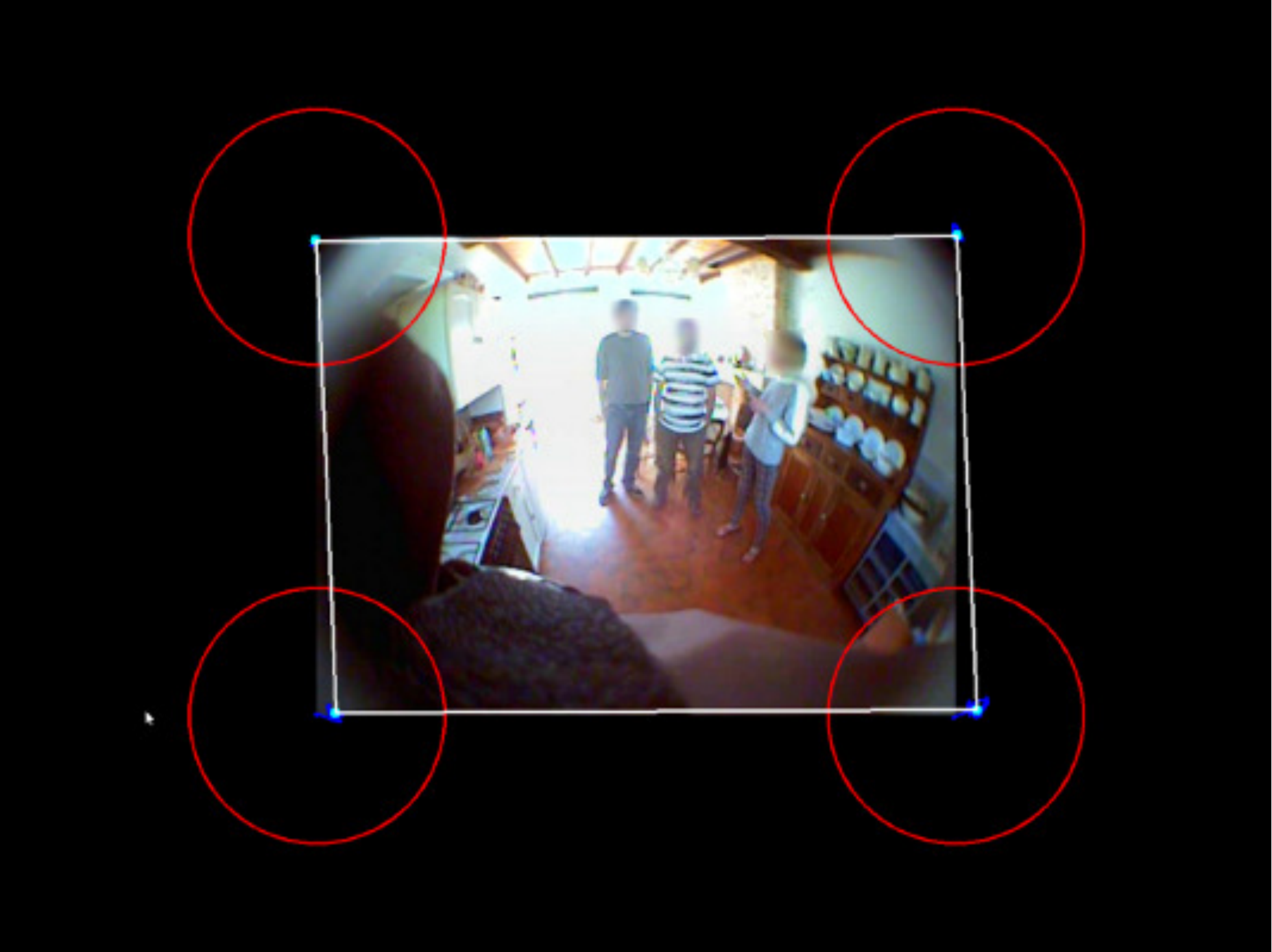}}
  \subfloat[Corner trajectories while the person moves to the left. ]{\label{Fig.Corners.Dynamic}	\includegraphics[width=4.1cm]{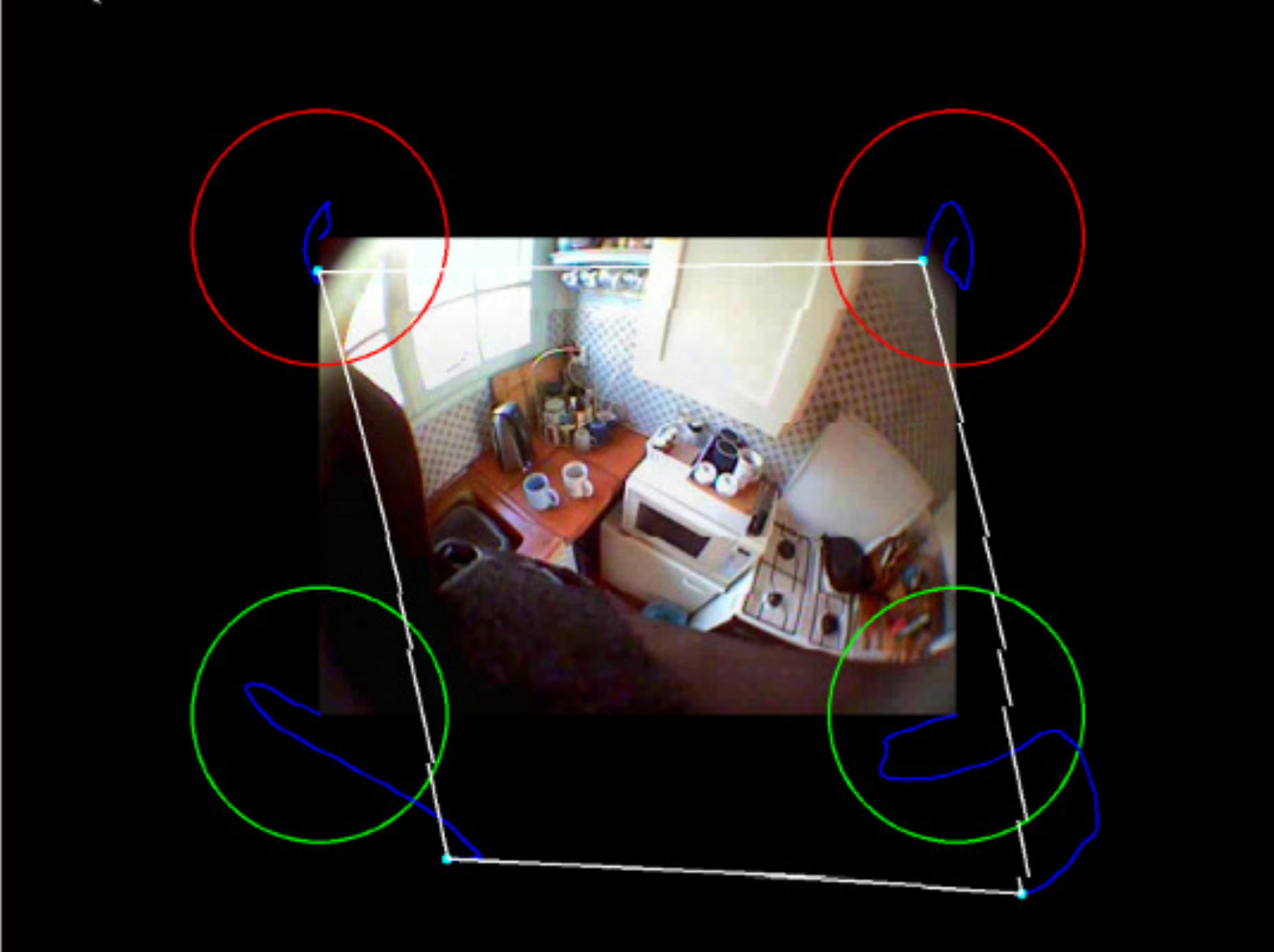}}
  \caption{Example of corners trajectories. }
  \label{Fig.Corners}
\end{figure}

To this aim, we compute the trajectories of each corner using the global motion estimation previously presented. For each frame the distance between the initial and the current position of a corner is calculated. We denote by $w$ the image width and by $s$ a threshold on the frame overlap rate. A corner is considered as having reached an outbound position when it has at least once had a distance greater than $s*w$ from its initial position in the current segment. These boundaries are represented by red and green (when the corner has reached an outbound position) circles in figure~\ref{Fig.Corners}.

\noindent

\subsubsection{Segment definition}

\begin{figure}[h]
 \centering
  \includegraphics[width=2.72cm]{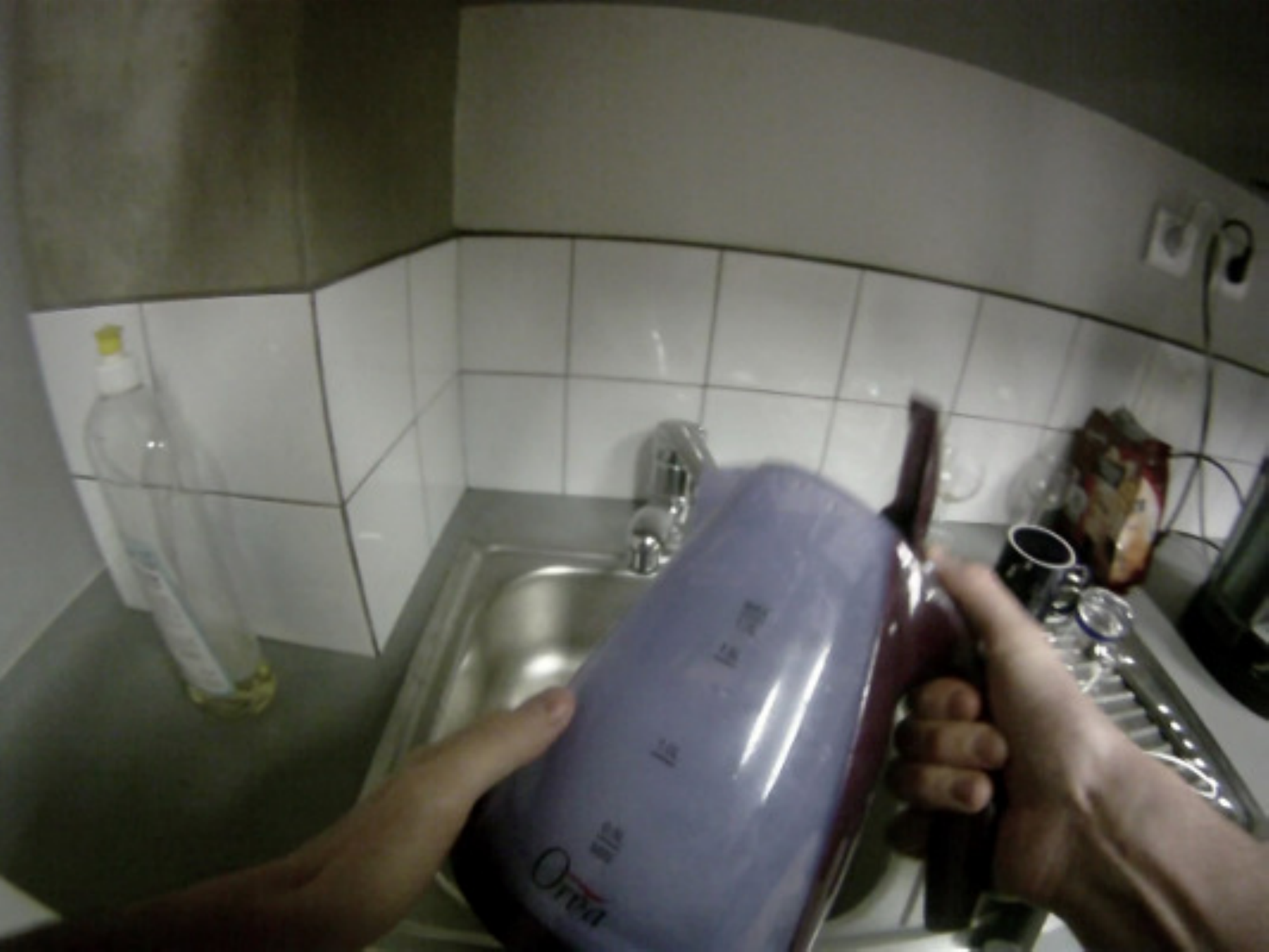}
  \includegraphics[width=2.72cm]{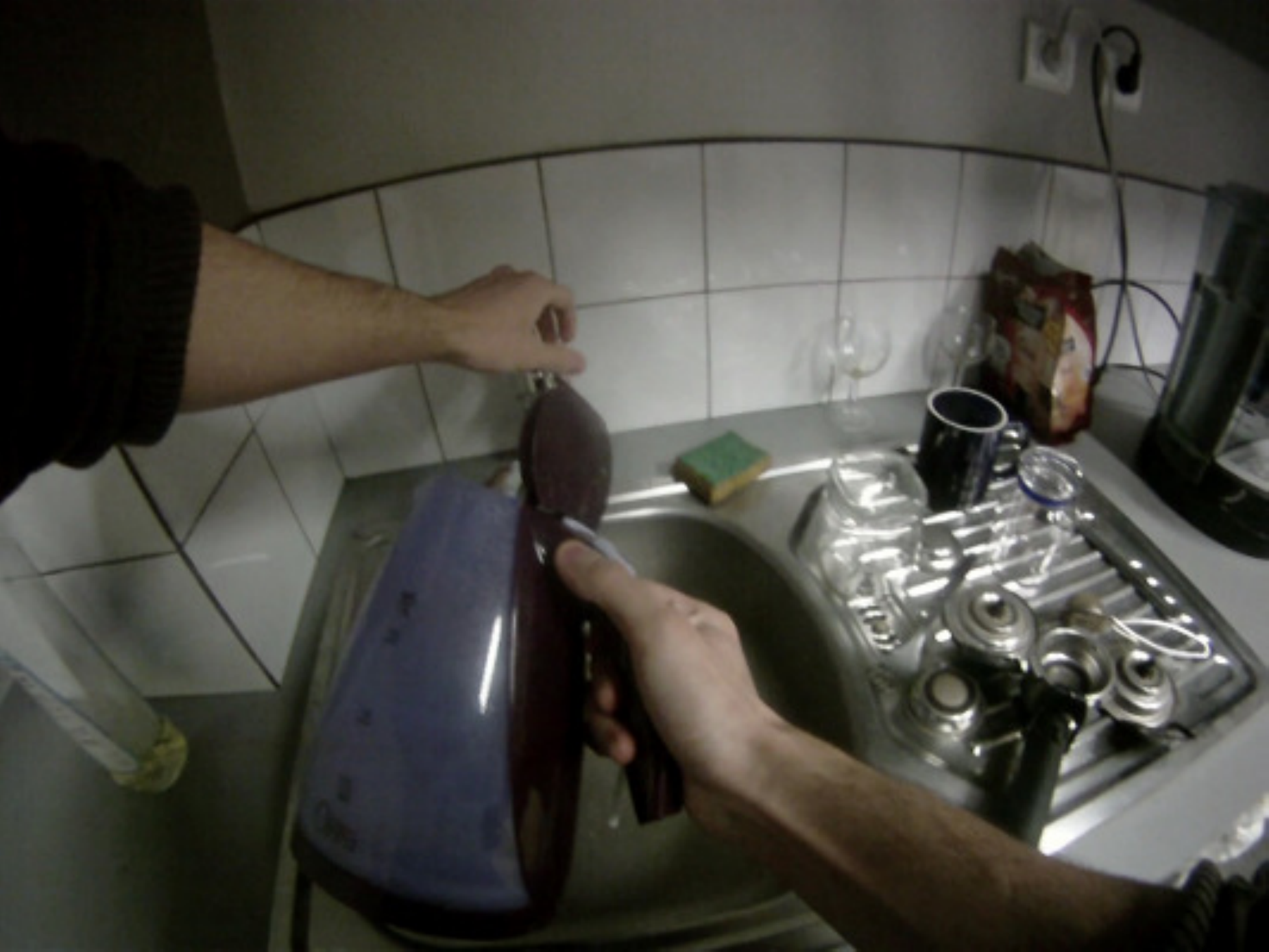}
  \includegraphics[width=2.72cm]{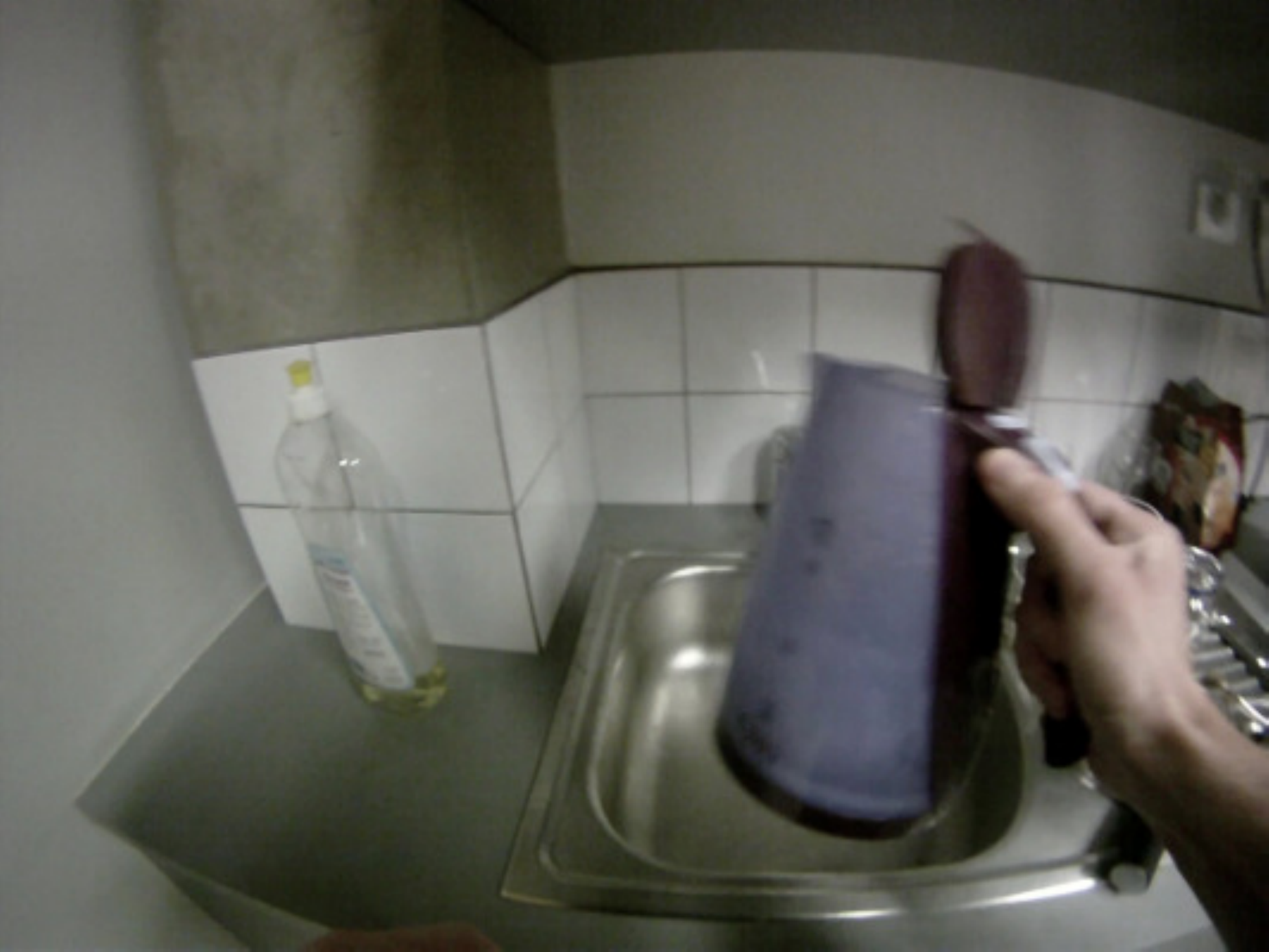}
  \caption{An example of key frame (center) with the beginning (left) and ending (right) frames of the segment. }
  \label{Fig.KeyFrames}
\end{figure}

Each segment aims to represent a single ``viewpoint''. This notion of viewpoint is clearly linked to the threshold~$s$, which defines the minimal proportion of an image which should be contained in all the frames of the segment. We define the following rules: a segment should contain a minimum of 5 frames and a maximum of 1000 frames, the end of the segment is the frame corresponding to the time when at least 3 corners have reached at least once an outbound position. The key frame is then chosen as the temporal center of the segment, see examples in figure~\ref{Fig.KeyFrames}.

\noindent

Hence the estimated motion model serves for two goals: i) estimated motion parameters are used for the computation of dynamic features in the global description space and ii) the key frames extracted form motion-segmented ``view points'' are the basis for extraction of spatial features. We will now focus on the definition of these two subspaces and the design of the global description space. 

\Section{Design of the description space}
\label{Section.DescriptionSpace}
The motion is one of the most important information in the videos studied. It represents the movements of the person and in a longer term history characterizes whether the action being done is dynamic or rather static.

\SubSection{Dynamic descriptors}

\subsubsection{Instant motion}

The ego-motion is estimated by the global motion analysis presented in section~\ref{Section.MotionAnalysis}. The parameters $a_{1}$ and $a_{4}$ are the translation parameters. We limit our analysis to these parameters, as in the case of wearable cameras, they express the dynamics of the behavior the best, and pure affine deformation without any translation is practically never observed. A histogram of the energy of each translation parameter $H_{tpe}$ is built according to Eq~\ref{Eq.Htpe}, defining a step $s_{h}$ and using a log scale. This histogram characterizes the instant motion. It is computed for each frame and then averaged over all the frames of a segment.

\begin{equation}	
\label{Eq.Htpe} \nonumber
\begin{array}{r r l l}
\multicolumn{4}{l}{H_{tpe}[i]+=1 \text{ if }} \\
 & log(a^2) & <  i \times s_{h}  & \text{ for } i = 1 \\
(i-1) \times s_{h} & \leq log(a^2) & < i \times s_{h} & \text{ for } i = 2 \text{..} N_{e}-1\\
i \times s_{h} & \leq  log(a^2) &   & \text{ for } i = N_{e} 
\end{array} 
\end{equation}
Eq.~\ref{Eq.Htpe}: Translation parameter histogram, a is either $a_{1}$ or $a_{4}$.\\

We denote $H_{tpe}(x)$ the histogram of the log energy of horizontal translation, and $H_{tpe}(y)$ the histogram of the energy of vertical translation observed in image plane. The number of bins is chosen the same $N_{e} = 5$, the threshold $s_{h}$ is chosen in such a way that the last bin corresponds to the translation of the image width or height respectively. 

\subsubsection{Motion history}
Another element to distinguish static and dynamic activities is the motion history. On the contrary to the instant motion we design it to characterize long-term dynamic activities, such as walking ahead, vacuum cleaning, etc... The estimation of this is done by computing a ``cut histogram"~$H_{c}$. We design it as a histogram of $i = 1-N_{c}$ bins. Each bin $H(i)$ contains the number of cuts (according to the motion based segmentation presented in section~\ref{Section.MotionAnalysis}) that happened in the last $2^{i}$ frames. The number of bins $N_{c}$ is defined as 8 in our experiments providing a history horizon of 256 frames, which represent almost 9 seconds for our 30 fps videos.

\SubSection{Static descriptors}
Static descriptors are computed on the extracted key frames representing each segment. In this choice we seek for the global descriptors which characterize the color of frames still preserving some spatial information. The MPEG-7 Colour Layout Descriptor (CLD) proved to be a good compromise for both \cite{Quenot2008}. It is computed on each key frame and the classical choice \cite{Sikora2002} of selecting  $6$ parameters for the luminance and $3$ for each chrominance was adopted.

\SubSection{Audio descriptors}
The particularity of our contribution in the design of a description space consists in the use of low-level audio descriptors. Indeed, in the home environment, with ambient TV audio track, noise produced by different objects the patient is manipulating, his conversations with the persons, are good indicators of activity and its location.\\
In order to characterize the audio environment, different sets of features are extracted. Each set is characteristic of a particular sound: speech, music, noise and silence \cite{Pinquier2006}. Energy is used for silence detection. 4 Hz energy modulation and entropy modulation give voicing information, being specific to the presence of speech.  The number of segments per second and the segment duration, resulting from a ``Forward-Backward'' divergence algorithm \cite{Andre1988}, are used to find harmonic sound, like music. Spectral coefficients are proposed to detect noise: percussion and periodic sounds (examples: footstep, flowing water, vacuum cleaner, etc.). 

\SubSection{Description space}
Hence for description of the content recorded with wearable cameras we designed three descriptors subspaces: the ``dynamic'' subspace has $18$ dimensions, and contains the descriptors D=($H_{tpe}(x)$,$H_{tpe}(y)$,$H_{c}$); the ``static'' subspace contains $l=12$ CLD coefficient C=($c_{1}$, ... ,$c_{l}$); the ``audio'' subspace contains $k=5$ audio descriptors p=($p_{1}$, ... ,$p_{k}$).\\
We design the global description space in an ``early fusion'' manner concatenating all descriptors in an observation vector $o$ in $R^{n}$ space with $n=35$ dimensions when all descriptors are used. Thus designed the description space is inhomogeneous. We also study the completeness and redundancy of this space in a pure experimental way with regard to the indexing of activities in Section 6. 

\Section{Design of an HMM structure}
\label{Section.HMM}

\begin{figure}[h]
 \centering
  \includegraphics[width=5.5cm,height=7cm]{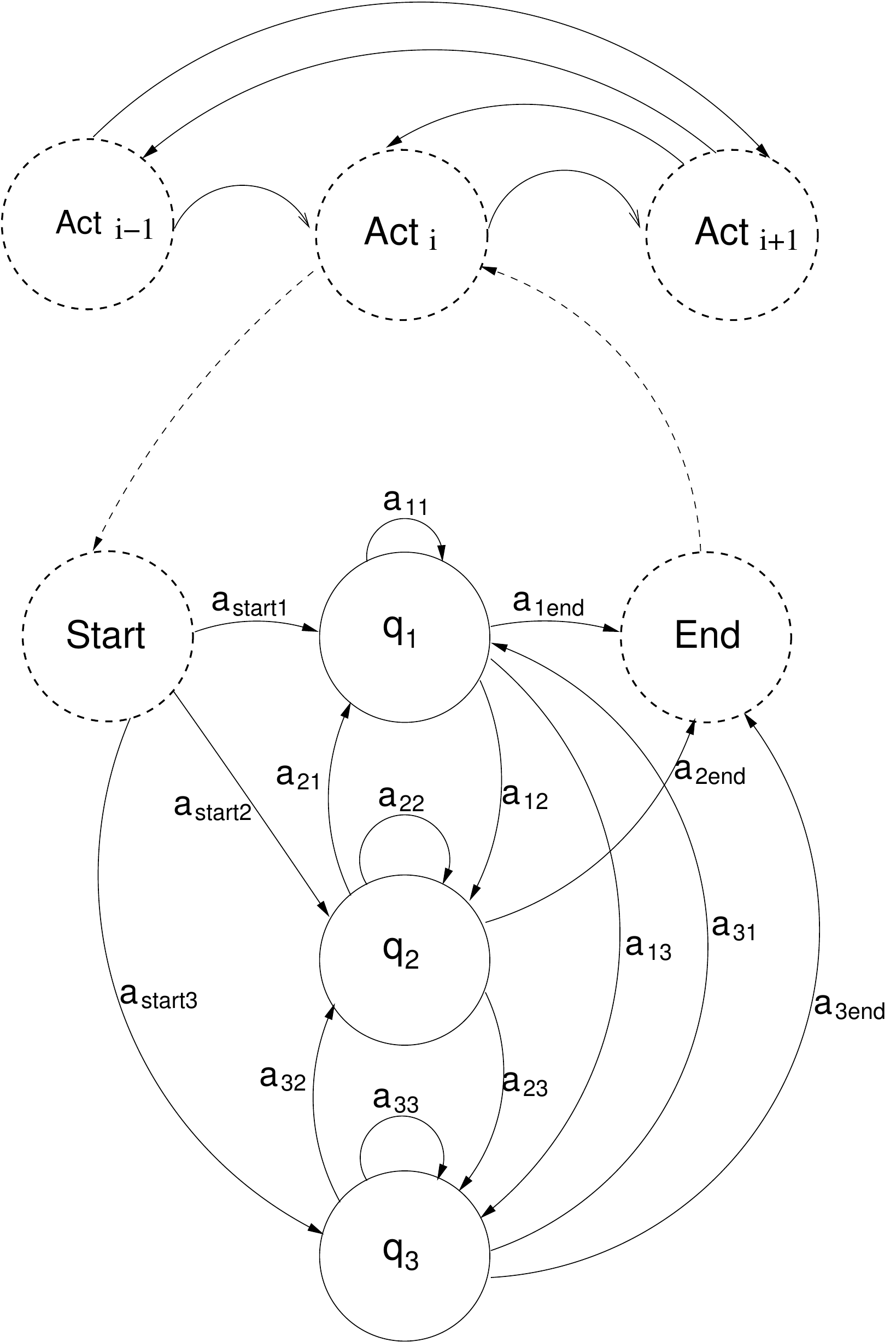}
  \caption{The HMM structure. }
  \label{Fig.HMM}
\end{figure}

If we consider our problem of recognition of daily activities in the video in a simplistic manner, we can draw an equivalence between an activity and a hidden state of an HMM. The connectivity of the HMM then can be defined by the spatial constraints of patient's environment. The easiest way is to design a fully connected HMM and train the inherent state-transition probabilities form the labeled data. Unfortunately, the ADL we consider are very much heterogeneous and often very complex. Hence we propose a two-level HMM. The activities meaningful for medical practitioners are encoded in the top-level HMM. It contains the transitions between ``semantic'' activities. A bottom level HHM models an activity with $m$ non-semantic states. This parameter $m$ is defined as 3, 5 or 7 in our experiments. The overall structure of the HMM is presented in figure~\ref{Fig.HMM}, with $3$ states at the bottom level. Dashed circled states are non emitting states. The HMMs are built using the HTK library \footnote{HTK Web-Site: \texttt{ http://htk.eng.cam.ac.uk}}.

\noindent

\SubSection{Top level HMM}

In this work, the actions of interest are the ADLs ``making coffee'', ``making tea'', ``washing the dishes'', ``discussing'', ``reading'' and another activity for all the rest which is not relevant to the ADLs of interest named ``NR". The top level HMM represents the relations between these actions. In this work no constraints were specified over the transitions between these activities since such restrictions did not apply in our application, hence we design the top level as a fully connected HMM.

\SubSection{Bottom level HMM}

Most of the activities defined in the above section are complex and could not easily be modeled by one state. For each activity in the top level HMM a bottom level HMM is defined. The bottom level HMM is composed of $m$ non semantic states. Each state models the observation vector~$o$, see section~\ref{Section.DescriptionSpace}, by a Gaussian Mixture Model (GMM). The GMM and the transitions matrix of all the bottom level HMM are learned using the classical Baum Welsh algorithm with labeled data corresponding to each activity.

\Section{Experiments}
\label{Section.Experiments}

Today, no rich corpus of data from wearable video settings has been publicly released. We can reference the dataset \cite{kitchenCMU} for a very limited task of behavior in the kitchen, where subjects are cooking different recipes. The only corpus recorded for ADL is ours. This corpus of 28 hours of videos contains heterogeneous activities, for this paper we used only a part of it to ensure multiple occurrences of activities for the supervised learning. The dataset used for this experiment comprises 6 videos shot in the same laboratory environment, containing a total of 81435 frames which represent more than 45 minutes. In these videos 6 activities of interest appear: ``working on computer'', ``reading'', ``making tea'', ``making coffee'', ``washing the dishes'', ``discussing'' and we added a reject class called ``NR". It represents all the moments which do not contain any of the activities of interest. The activities of interest are the ones present in the survey the doctors were using until now. We use a cross validation, the HMMs models of activities were learnt on all but one video and tested on this excluded video. We will first discuss the influence of the segmentation parameters and the choice of the description space and finally analyze our results on activities recognition.

\SubSection{Segmentation analysis}

\noindent
\begin{figure}
 \centering
  \includegraphics[width=8cm,height=4.9cm]{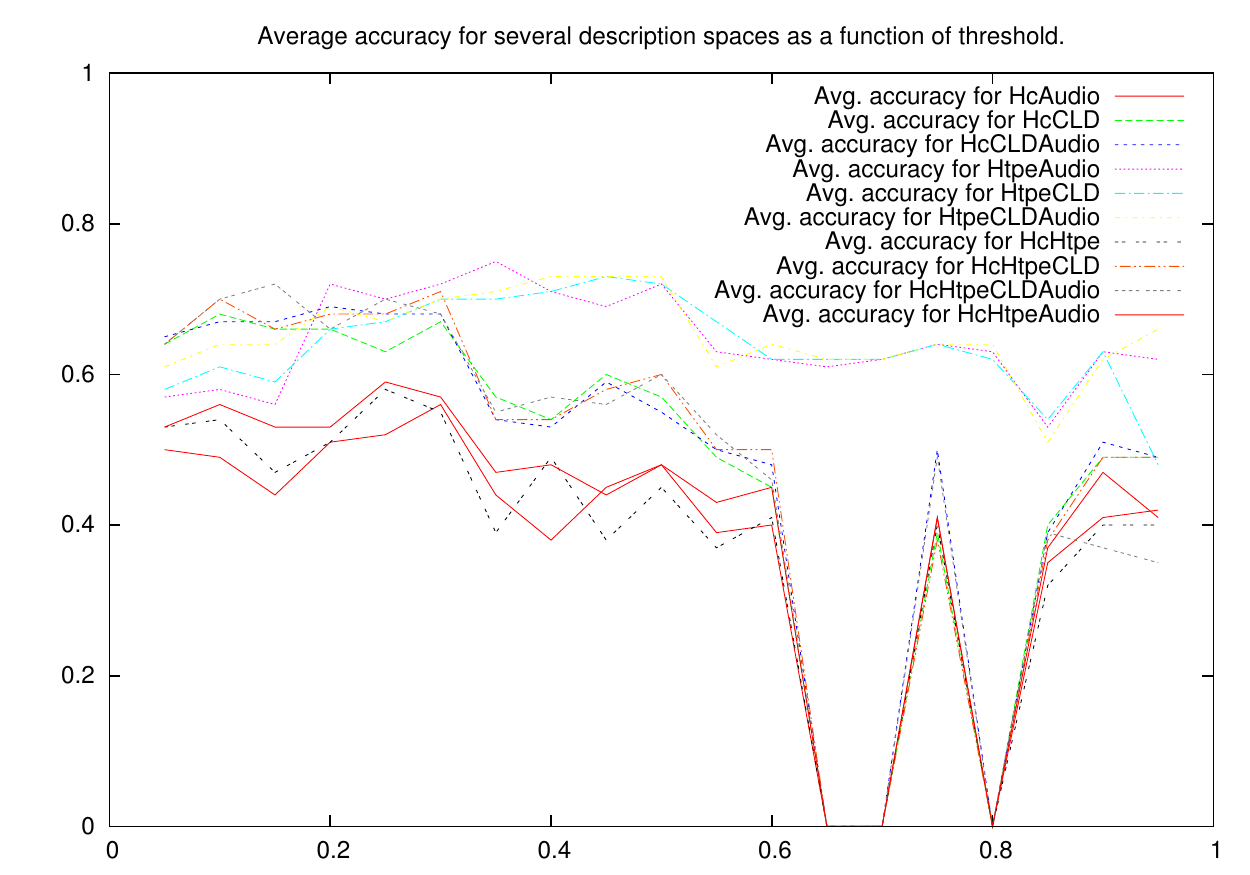}
  \caption{Description space choice and segmentation threshold influence over accuracy. }
  \label{Fig.DescSpace}
\end{figure}

The influence of the segmentation threshold is not as significant as we expected but figure~\ref{Fig.DescSpace} shows that the accuracy starts to decrease for threshold values higher than 0.3. Indeed, the higher the threshold is, the probability of having a segment containing different activities increases. The activity ``making coffee'' and ``washing the dishes'' may follow each other in a short time. Moreover, the higher the threshold is the less data are available for the HMM training. This explains the fall to zero in some curves when there is not enough data to train the HMM.

\SubSection{Study of the description space}

The description space is defined as one of the possible combinations of the descriptors presented in section~\ref{Section.DescriptionSpace}. Figure~\ref{Fig.DescSpace} presents the average accuracy for different combinations of descriptors as a function of the segmentation threshold parameter. The performances of the $H_{tpe}CLDAudio$ (yellow) and $H_{tpe}Audio$ (pink) descriptor indicate the positive contribution of the audio descriptor.\\
The CLD descriptor seems to improve the results for low segmentation thresholds, all the six best description space configurations in figure 8, for threshold less than 0.1, contains the CLD descriptor. This is rather normal since larger the segment is, less the CLD of the key frame will be meaningful regarding the content of the segment.\\
The full description space $H_{c}H_{tpe}CLDAudio$ (gray dashed curve) performs really well for the 0.1 threshold. Being more complex this description space also needs more training data, therefore with higher thresholds the performance falls.

\SubSection{Activity recognition}

\subsubsection{HMM analysis}

\noindent
\begin{figure}
 \centering
  \includegraphics[width=8.6cm,height=5cm]{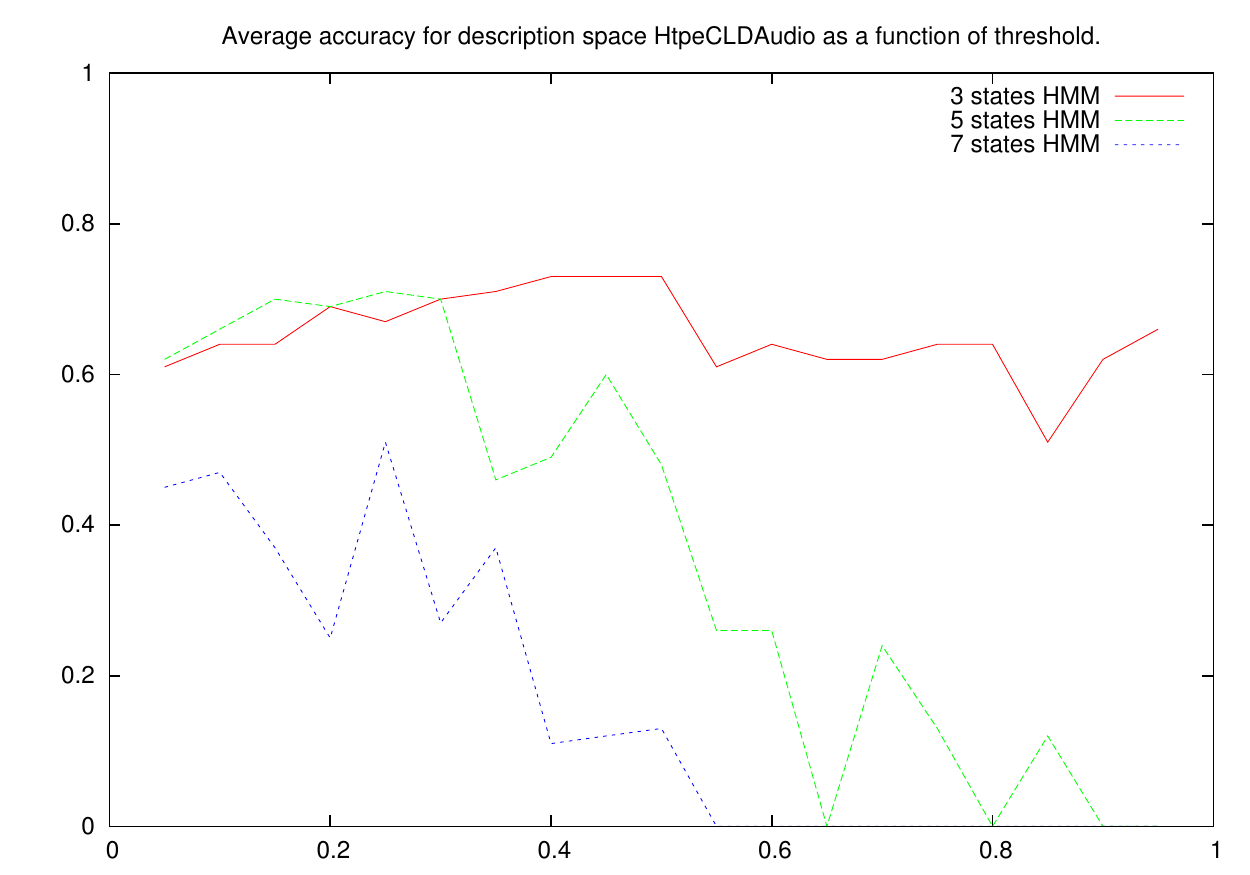}
  \caption{Analysis of number of states in HMMs. }
  \label{Fig.NumberStatesHMMs}
\end{figure}

In our experiments we have found that with a higher threshold less data become available for the HMM training which is a significant issue. Therefore, with less data available only the configuration with 3-states still performs well, see figure~\ref{Fig.NumberStatesHMMs}. The 7-states configuration falls to zero for threshold higher than 0.5, and the 5-states configuration is quite unstable for threshold values higher than 0.65.

\subsubsection{Activities recognition}

\begin{figure}
 \centering
  \includegraphics[width=8.6cm,height=5cm]{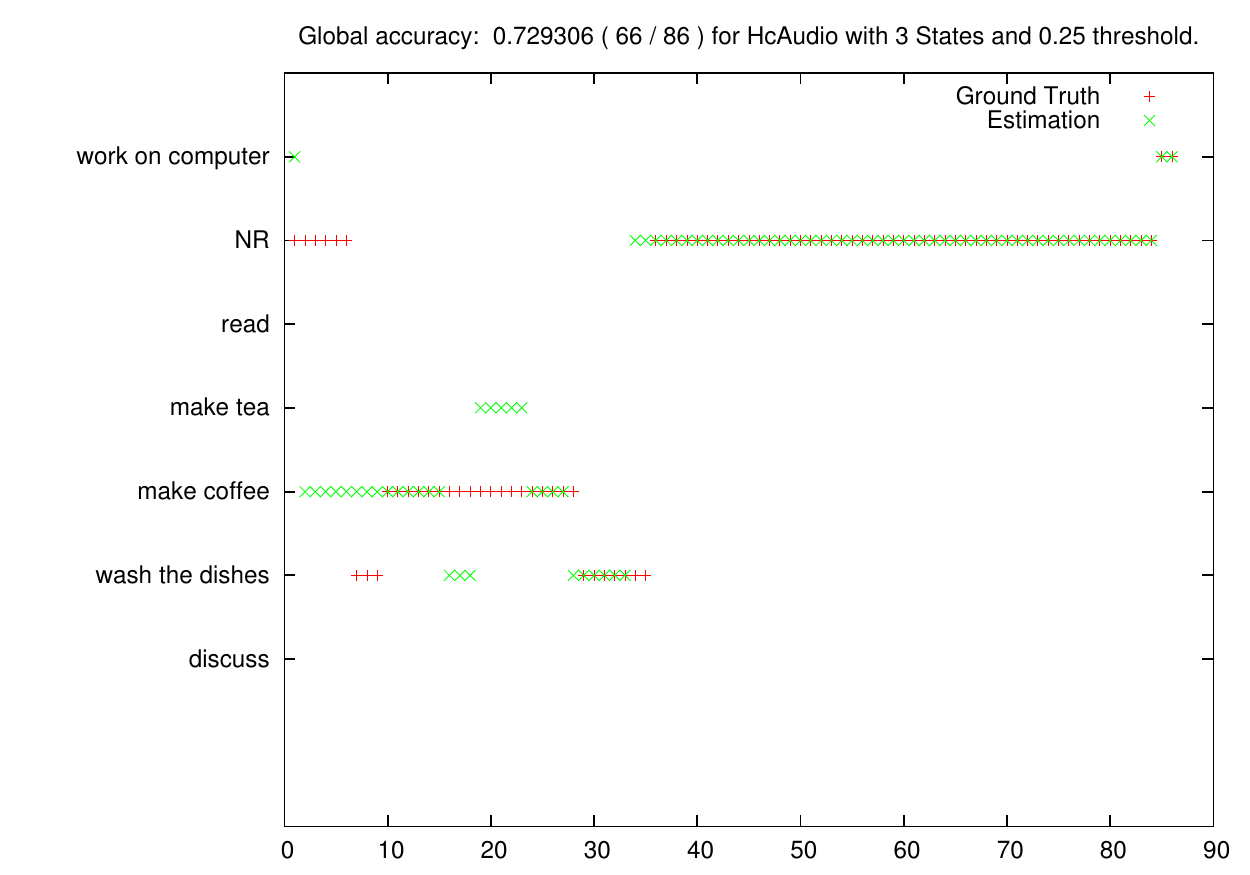}
  \caption{Results of the analysis. }
  \label{Fig.ResultsAnalysis}
\end{figure}

In order to evaluate the ADL recognition we have chosen one of the average recognition results presented in figure~\ref{Fig.ResultsAnalysis}. The ``reading'' and ``discussing'' activities are not present and are not detected for this video. The main confusions result in the activities ``making coffee'' and ``washing the dishes". These activities are similar in terms of environment as well as motion and audio characteristics. The activity ``working on computer'' is hard to define with the description space $H_{c}H_{tpe}Audio$ presented, therefore some misdetections appear.
 
\noindent

\Section{Conclusions and perspectives}
\label{Section.ConclusionsAndPerspectives}

In this paper, we have presented a method for indexing video sequences acquired from a wearable camera. We have proposed an original approach to segment the video into temporally consistent viewpoints, thanks to apparent motion analysis. This segmentation has been used to define new motion descriptors. Motion, color and audio features have been used as multimodal observation in a hierarchical Hidden Markov Model, applied to the task of recognizing a set of activities of interest.\\
The confusion amongst activities show that the global descriptors may be close for different activities. Since the person does not interact with the same objects for different activities, our future work will be to detect the objects of interest and the eventual interaction of the person with them to better characterize the ADLs. Despite the experimental data set has not been very large yet, this research gave a ``proof of concept'' and opens tremendous perspectives for our future work.

\Section{Acknowledgments}

This work is supported by a grant from Agence Nationale de la Recherche with reference ANR-09-BLAN-0165-02, within the IMMED project.


\bibliographystyle{latex8}
\bibliography{latex8}

\end{document}